\documentclass[10pt,conference]{IEEEtran}
\IEEEoverridecommandlockouts
\usepackage{cite}
\usepackage{amsmath,amssymb,amsfonts}
\usepackage{algorithmic}
\usepackage{graphicx}
\usepackage{textcomp}
\usepackage{xcolor}
\usepackage{enumitem}
\usepackage{subcaption}

\def\BibTeX{{\rm B\kern-.05em{\sc i\kern-.025em b}\kern-.08em
    T\kern-.1667em\lower.7ex\hbox{E}\kern-.125emX}}
\begin{document}

\title{\textbf{COMET}: \textbf{C}ombinatorial \textbf{O}ptimization for \textbf{M}ultiplex \textbf{E}diting \textbf{T}argets Via Constraint-Preserving QAOA}

\author{
\IEEEauthorblockN{Priyansh Singhal}
\IEEEauthorblockA{\textit{Independent Researcher} \\
\textit{India}\\
priyansh.s21@iiits.in}
\and
\IEEEauthorblockN{Sumit Maheshwari\textsuperscript{\dag}}
\IEEEauthorblockA{\textit{Microsoft}\\
Boston, USA\\
sumit.maheshwari@microsoft.com}
\and
\IEEEauthorblockN{Piyush Joshi}
\IEEEauthorblockA{\textit{Department of Computer Science \& Engineering} \\
\textit{IIIT Sri City, India}\\
piyush.j@iiits.in}

\thanks{\textsuperscript{\dag}Work done in personal capacity; does not reflect the views of Microsoft.}
}

\maketitle

\begin{abstract}
Multiplex CRISPR-Cas9 gene editing requires selecting one guide RNA per target gene subject to cross-gene interactions: a constrained combinatorial problem that can be formulated as a Quadratic Unconstrained Binary Optimization (QUBO) and solved via the Quantum Approximate Optimization Algorithm (QAOA). The one-hot per-gene constraint is conventionally enforced by adding quadratic penalty terms to the cost Hamiltonian, but penalty coefficient selection is heuristic and penalties amplify hardware noise. An alternative is to enforce the constraint structurally via the XY-mixer, which preserves feasibility by construction. We present COMET, a systematic comparison of penalty-based and XY-mixer QAOA on a three-gene, twelve-qubit multiplex editing instance targeting the immune-checkpoint genes \textit{PDCD1}, \textit{LAG3}, and \textit{HAVCR2}. In simulation, the XY-mixer exceeds $95\%$ probability of the optimum by QAOA depth $p{=}3$, while three penalty variants spanning an order of magnitude in penalty coefficient remain below $6\%$ at every depth. On IBM's \texttt{ibm\_kingston} (Heron r2) processor, the XY-mixer's simulator--hardware energy gap stays within $|0.8|$ across all depths, while the worst-tuned penalty variant's gap reaches $+53.9$. We provide an honest account of where the structural guarantee partially breaks under gate-level noise. The twelve-qubit instance is classically trivial; our contribution is a methodological comparison of constraint-enforcement strategies in a biologically motivated domain, with real-hardware validation.
\end{abstract}

\section{Introduction}
\label{sec:intro}

Multiplex CRISPR-Cas9 \cite{doudna2014new, mccarty2020multiplexed} editing which is simultaneous disruption of multiple genomic loci in a single experiment has emerged as an essential capability in cell and gene therapies, particularly in cancer immunotherapy \cite{liu2017crispr}, where concurrent knockout of redundant immune-checkpoint genes can amplify anti-tumor T-cell activity~\cite{cong2013multiplex, mali2013rna, ciraolo2022simultaneous} A practical bottleneck in such experiments is the selection of guide RNAs (gRNAs): each gene admits many candidate gRNAs, each scored along on-target efficacy and off-target risk~\cite{doench2016optimized}, and for multiplex experiments these choices interact across genes. A gRNA that is individually strong may be a poor partner to the strong picks for other genes in the panel, owing to synergistic toxicity, resource competition, or beneficial synergy. The selection problem is therefore combinatorial over the joint space of per-gene candidates.

We demonstrate this concretely in a three-gene instance (\textit{PDCD1}~\cite{ishida1992induced}, \textit{LAG3}~\cite{chocarro2021understanding}, \textit{HAVCR2}~\cite{monney2002th1}) with four candidates per gene. The biologist's natural heuristic (pick the most potent candidate per gene) produces the \emph{worst} of all 64 feasible selections at rank 64/64, because the three individually-best candidates share the strongest positive pairwise couplings. The true optimum, which differs at every gene from the greedy pick, cannot be obtained by any per-gene ranking. This ``greedy trap'' motivates formulation of multiplex gRNA selection as a constrained combinatorial optimization problem.

Such problems map naturally to Quadratic Unconstrained Binary Optimization (QUBO), and from there to the cost Hamiltonian of the Quantum Approximate Optimization Algorithm (QAOA)~\cite{farhi2014quantum}. However, QAOA's canonical formulation cannot directly represent the one-hot constraint (one gRNA per gene): constraints must be absorbed into the cost Hamiltonian as quadratic penalty terms weighted by a coefficient $P$. This strategy has well-documented weaknesses. The penalty coefficient must be tuned manually or heuristically, and its choice materially affects the optimization landscape~\cite{roch2023effect}. More fundamentally, penalties inflate the cost Hamiltonian's operator norm, which both distorts optimization and amplifies hardware noise during expectation-value estimation.

An alternative, introduced in the Quantum Alternating Operator Ansatz framework~\cite{hadfield2019quantum}, is to enforce constraints structurally via a mixer Hamiltonian that preserves feasibility by construction. For one-hot constraints, the \emph{XY-mixer}~\cite{rieffel2020xy} commutes with the per-group Hamming-weight operator; when initialized from a one-hot feasible state, evolution under the XY-mixer remains confined to the feasible subspace, eliminating the need for penalty terms in the cost Hamiltonian. This shifts constraint enforcement from the objective (soft) to the ansatz (hard).

In this paper we present \textbf{COMET} (Combinatorial Optimization for Multiplex Editing Targets), a proof-of-concept comparison of penalty-based and XY-mixer QAOA on multiplex gRNA selection, validated on both noiseless simulator and IBM's \texttt{ibm\_kingston} (Heron r2) quantum processor with systematic error-mitigation analysis.

Our contributions are as follows.
\begin{itemize}[leftmargin=*,topsep=2pt,itemsep=1pt]
    \item We formulate multiplex gRNA selection as a constrained QUBO with linear on/off-target costs and pairwise cross-gene interactions, and exhibit a three-gene instance in which greedy per-gene selection is catastrophically suboptimal.
    \item We compare four QAOA variants (three penalty coefficients spanning an order of magnitude, and the XY-mixer) across depths $p{=}1$--$5$ in simulation, showing that the XY-mixer reaches $>95\%$ probability of the optimum while all penalty variants remain below $6\%$.
    \item We execute all four variants on \texttt{ibm\_kingston} with systematic error-mitigation comparison, reporting energy, feasibility, and optimum-sampling metrics with repetition-based confidence bands.
    \item We provide an honest account of where structural constraint enforcement succeeds and where it partially fails on present hardware, including feasibility degradation under gate noise and zero-noise extrapolation overcorrection at depth.
\end{itemize}

We emphasize that the 12-qubit problem size is classically trivial, and we make no claim of quantum advantage. Our contribution is methodological: a systematic comparison of constraint-enforcement strategies in a biologically motivated application, with real-hardware execution rather than simulator-only analysis.

The remainder of the paper is organized as follows. Section~\ref{sec:related} surveys related work. Sections~\ref{sec:background}--\ref{sec:methods} review the relevant QAOA theory, formalize the gRNA selection problem, and describe our methodology. Sections~\ref{sec:sim_results} and \ref{sec:hw_results} report simulator and hardware results. Section~\ref{sec:discussion} discusses implications and limitations, and Section~\ref{sec:conclusion} concludes.
\section{Related Work}
\label{sec:related}

\subsection{QAOA and Constraint-Preserving Mixers}
QAOA~\cite{farhi2014quantum} has become a standard near-term approach to combinatorial optimization, with a substantial literature on variants and performance characterization~\cite{blekos2024review}. The Quantum Alternating Operator Ansatz framework~\cite{hadfield2019quantum} generalizes the original QAOA mixer to families of unitaries that preserve problem-specific feasibility, and provides the theoretical basis for our XY-mixer construction. Technical foundations for the XY-mixer on one-hot constrained problems, including fermionic-simulation depth analyses for ring and complete-graph variants are developed by Wang et al.~\cite{wang1904xy}, with efficient Trotterization schemes for general constraint subspaces given by Fuchs et al.~\cite{fuchs2022constraint}. Analytical frameworks for the approximation performance of structure-preserving ansätze are given by Hadfield and Hogg~\cite{hadfield2023analytical}. Recent work has scaled warm-started XY-QAOA to a 144-qubit IBM processor on Max-$k$-Cut and TSP benchmarks~\cite{bucher2026constrained}, and introduced adaptive Hamming-weight operators as a distinct constraint-aware construction validated on portfolio optimization and high-energy physics benchmarks~\cite{hao2026constraint}. Our contribution is complementary: we apply the XY-mixer construction to a new application domain and provide systematic sim--hardware comparison with honest accounting of where the structural guarantee breaks under noise.

\subsection{QUBO Formulations and Penalty Methods}
Ising and QUBO formulations of NP-hard problems are surveyed in Lucas~\cite{lucas2014ising}, with a pedagogical tutorial by Glover et al.~\cite{glover2018tutorial} covering penalty-based constraint encoding. Theoretical and empirical analyses of penalty coefficient selection and of its effect on the spectral gap of constrained Hamiltonians are given by Roch et al.~\cite{roch2023effect}, whose findings motivate our systematic penalty sweep.

\subsection{Quantum Computing for Biology and Bioinformatics}
Quantum-algorithmic approaches in computational biology have concentrated primarily on protein folding, molecular docking, and genome assembly. Variants of QAOA have been applied to protein-ligand docking~\cite{ding2024molecular}, and de novo genome assembly has been formulated as a QUBO and solved on both D-Wave annealers and gate-based QAOA simulators~\cite{boev2021genome}. A recent review of quantum computing for genomics~\cite{maurizio2025quantum} surveys QAOA and VQE applications across the field but does not identify any prior work applying quantum optimization to CRISPR gRNA selection. To our knowledge, the only quantum-adjacent work on gRNA is the use of quantum-chemical descriptors as classical ML features for sgRNA efficiency prediction~\cite{noshay2023quantum}, categorically distinct from our approach, which applies quantum circuit-based optimization to the combinatorial selection problem itself.

Zero-noise extrapolation~\cite{temme2017error,li2017efficient}, its digital gate-folding variant~\cite{giurgica2020digital}, and associated best-practice analyses~\cite{majumdar2023best} form the error-mitigation basis of our hardware protocol, which we access through the Qiskit Runtime V2 primitives~\cite{javadi2024quantum}. Dynamical decoupling with the XY4 sequence~\cite{viola1999dynamical} is benchmarked on IBM superconducting qubits by Ezzell et al.~\cite{ezzell2023dynamical}, with QAOA-specific DD analyses provided by Ji and Polian~\cite{ji2024synergistic}. Utility-scale error-mitigated experiments on IBM Eagle processors~\cite{Kim2023Evidence} and empirical QAOA+ZNE studies on IBM Heron devices~\cite{Ribeiro2026Empirical} provide recent context for hardware-level QAOA. Theoretical bias analyses of ZNE~\cite{Mohammadipour2025Direct} frame our interpretation of the negative sim--hardware gaps observed on the XY-mixer at deep circuits.
\section{Background}
\label{sec:background}

\begin{figure*}[t]
\centering
\begin{subfigure}[b]{0.48\textwidth}
    \centering
    \includegraphics[width=\textwidth]{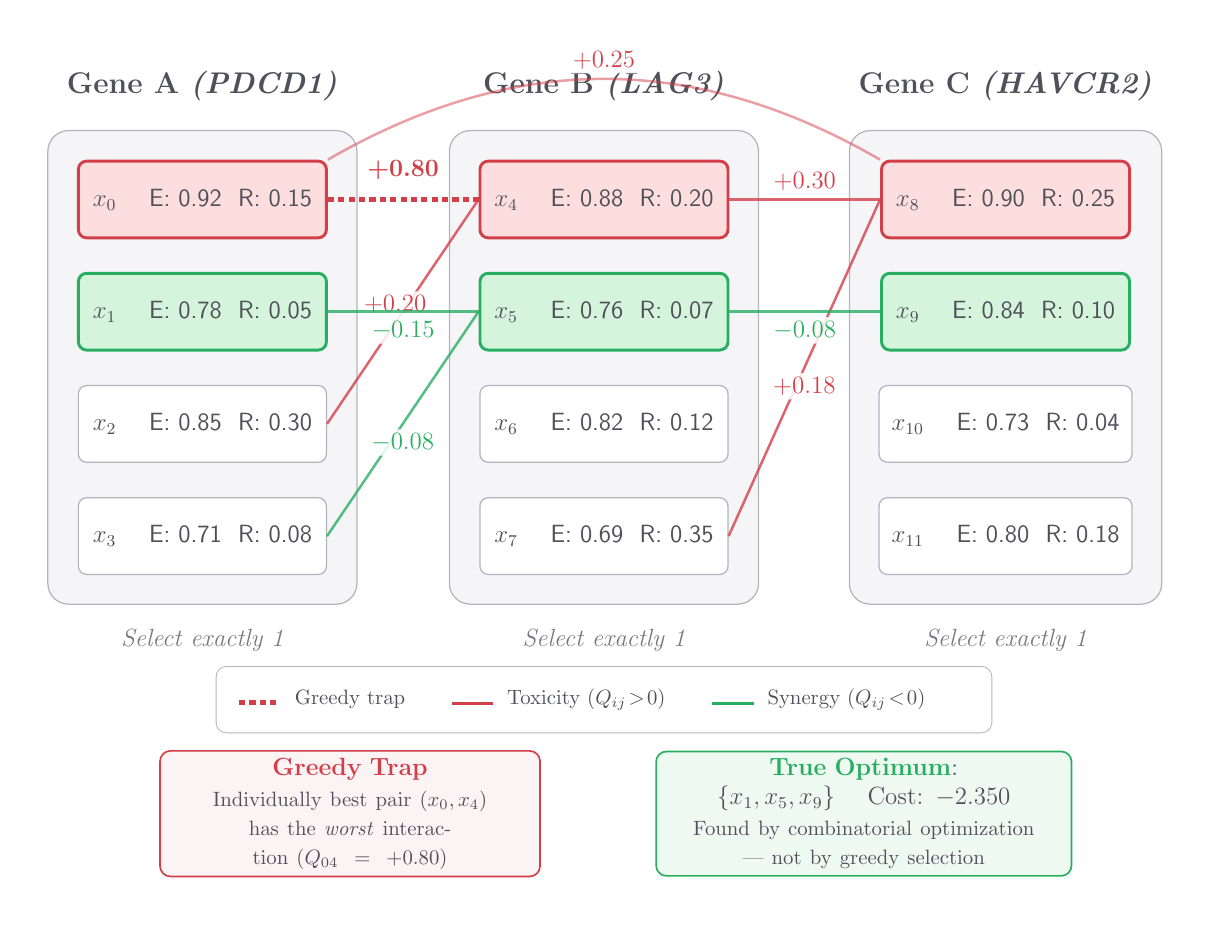}
    \caption{Three genes, four candidate gRNAs each, with linear costs $L_i$ and cross-gene pairwise interactions $Q_{ij}$.}
    \label{fig:problem_formulation}
\end{subfigure}
\hfill
\begin{subfigure}[b]{0.48\textwidth}
    \centering
    \includegraphics[width=\textwidth]{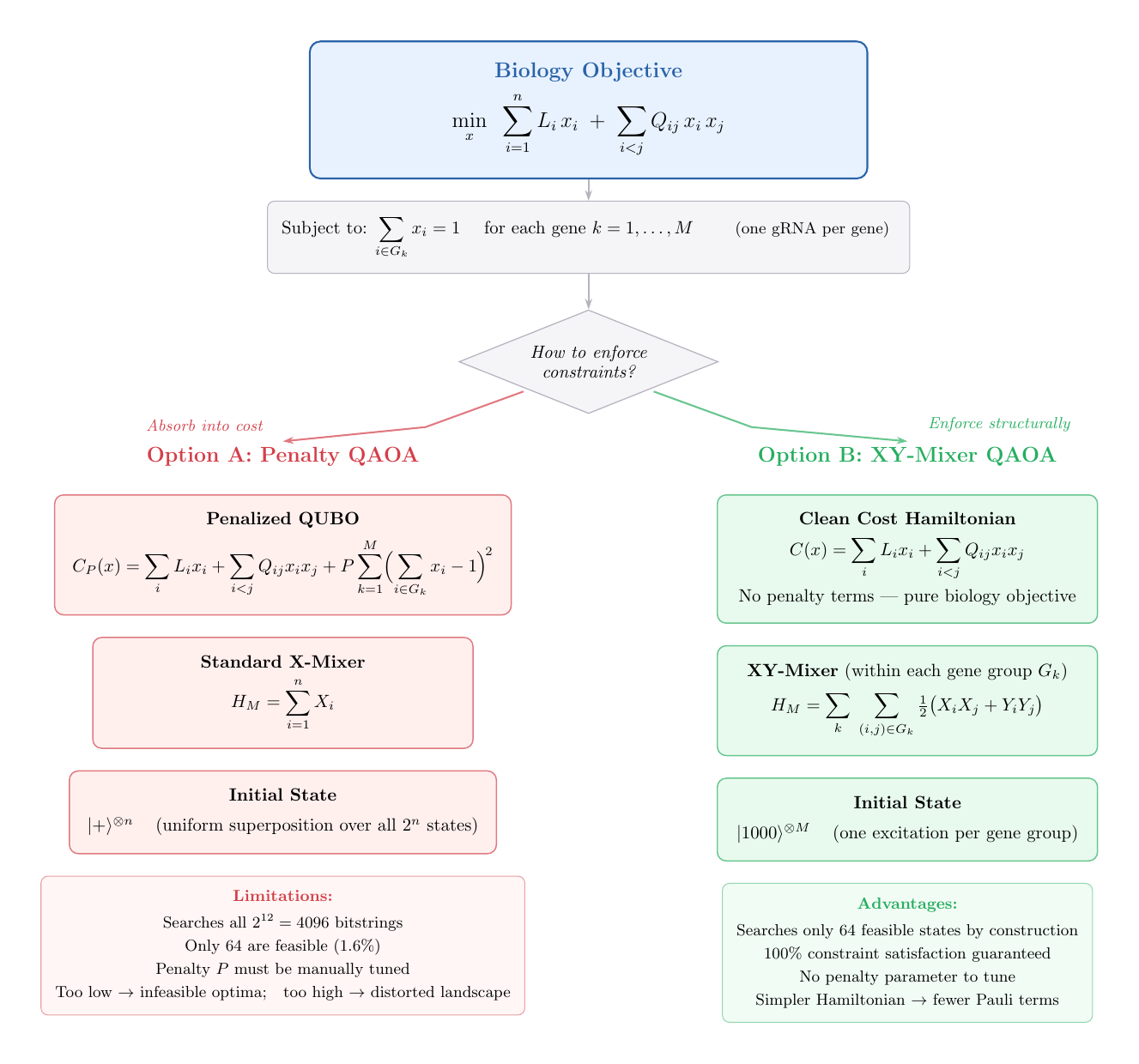}
    \caption{Two strategies for enforcing the one-hot constraint: penalty terms added to $H_C$ versus structural enforcement via the XY-mixer.}
    \label{fig:qubo_formulation}
\end{subfigure}
\caption{COMET problem structure and constraint-handling strategies.}
\label{fig:slot1}
\end{figure*}

\subsection{Multiplex CRISPR-Cas9 Editing}
CRISPR-Cas9 enables targeted genome editing by pairing a programmable guide RNA (gRNA) with the Cas9 nuclease, which then introduces a double-strand break at a sequence-specified genomic locus~\cite{cong2013multiplex,mali2013rna}. \emph{Multiplex} editing refers to simultaneous disruption of multiple loci using several gRNAs delivered together, a capability essential for therapies targeting redundant or synergistic gene pathways. In cancer immunotherapy, simultaneous knockout of the immune-checkpoint genes \textit{PDCD1} (PD-1), \textit{LAG3}, and \textit{HAVCR2} (TIM-3) in T cells has been demonstrated experimentally to amplify anti-tumor activity relative to single-gene disruption~\cite{ciraolo2022simultaneous}.

\subsection{gRNA Scoring and Selection}
Each gene offers many candidate gRNAs, and a candidate's quality is assessed along two axes: \emph{on-target efficacy}, the probability that the guide induces the intended cut, and \emph{off-target risk}, the propensity to cut unintended loci. Canonical scoring methods such as Rule Set 2 and the Cutting Frequency Determination (CFD) score provide quantitative estimates of both~\cite{doench2016optimized}. When editing a single gene, selection reduces to ranking candidates; multiplex editing, however, introduces cross-gene interactions, and the combinatorial selection problem over several genes can no longer be solved by independent per-gene ranking.

\subsection{QUBO and Ising Formulations}
A Quadratic Unconstrained Binary Optimization (QUBO) problem minimizes a quadratic form over binary variables $x \in \{0,1\}^n$:
\begin{equation}
\min_{x} \; \sum_{i} L_i x_i + \sum_{i<j} Q_{ij} x_i x_j,
\label{eq:qubo}
\end{equation}
and maps to an Ising Hamiltonian via $x_i = (1 - Z_i)/2$~\cite{lucas2014ising, glover2018tutorial}. Equality constraints $\sum_{i \in S} x_i = 1$ are conventionally enforced by adding penalty terms $P\bigl(\sum_{i \in S} x_i - 1\bigr)^2$ to the objective. The coefficient $P$ must exceed problem-dependent bounds to guarantee constraint satisfaction at the ground state; its choice materially affects the spectral gap and hence the optimization landscape~\cite{roch2023effect}.

\subsection{Quantum Approximate Optimization Algorithm}
QAOA~\cite{farhi2014quantum} prepares a variational state of depth $p$,
\begin{equation}
|\boldsymbol{\gamma}, \boldsymbol{\beta}\rangle = \prod_{\ell=1}^{p} e^{-i\beta_\ell H_M} e^{-i\gamma_\ell H_C} |\psi_0\rangle,
\label{eq:qaoa}
\end{equation}
alternating between evolution under a cost Hamiltonian $H_C$ (encoding the QUBO) and a mixer Hamiltonian $H_M$, starting from an initial state $|\psi_0\rangle$. The $2p$ angles $(\boldsymbol{\gamma}, \boldsymbol{\beta})$ are classically optimized to minimize $\langle H_C \rangle$. Performance is provably non-decreasing in $p$~\cite{farhi2014quantum}. In the standard formulation, $H_M = \sum_i X_i$ and $|\psi_0\rangle = |+\rangle^{\otimes n}$, and all constraints are absorbed into $H_C$ as penalties. See~\cite{blekos2024review} for a recent survey of variants.

\subsection{Constraint-Preserving Mixers}
The Quantum Alternating Operator Ansatz generalizes QAOA by permitting any mixer Hamiltonian, including those that preserve feasibility with respect to problem constraints~\cite{hadfield2019quantum}. For one-hot constraints, the \emph{XY-mixer}
\begin{equation}
H_M^{XY} = \sum_{S} \sum_{i<j \in S} \tfrac{1}{2}(X_i X_j + Y_i Y_j)
\label{eq:xy_mixer}
\end{equation}
commutes with each group's Hamming-weight operator: each term swaps $|01\rangle \leftrightarrow |10\rangle$ within a group while annihilating $|00\rangle$ and $|11\rangle$~\cite{wang1904xy}. When initialized from a one-hot feasible state, evolution under $H_M^{XY}$ remains confined to the feasible subspace by construction, eliminating the need for penalty terms in $H_C$.
\section{Problem Formulation}
\label{sec:formulation}

\subsection{gRNA Selection as Combinatorial Optimization}
We consider the selection of one gRNA per gene across $M=3$ immune-checkpoint genes (\textit{PDCD1}, \textit{LAG3}, \textit{HAVCR2}), with $N=4$ candidate gRNAs per gene, yielding $n = MN = 12$ binary variables $x_i \in \{0,1\}$. Variable $x_i$ is 1 if candidate $i$ is selected. The one-hot constraint
\begin{equation}
\sum_{i \in S_k} x_i = 1, \quad k = 1, \ldots, M,
\label{eq:onehot}
\end{equation}
holds for each gene group $S_k = \{kN, kN+1, \ldots, kN+N-1\}$. The feasible set contains $N^M = 64$ of the $2^{12} = 4096$ possible bitstrings (1.6\%).

\subsection{Cost Function}
Each candidate carries a linear cost $L_i = w_{\text{off}} \cdot \text{Off}_i - w_{\text{on}} \cdot \text{On}_i$, where $\text{On}_i$ and $\text{Off}_i$ are on-target and off-target scores and $w_{\text{on}} = w_{\text{off}} = 1$. More negative $L_i$ denotes a higher-quality individual candidate. Cross-gene pairwise interactions $Q_{ij}$ model combinatorial effects: positive $Q_{ij}$ captures synergistic toxicity or resource competition, while negative $Q_{ij}$ captures beneficial synergy. The resulting objective,
\begin{equation}
C(x) = \sum_{i=1}^{n} L_i x_i + \sum_{i < j} Q_{ij} x_i x_j,
\label{eq:cost}
\end{equation}
takes the QUBO form of Eq.~\ref{eq:qubo}. While the numerical values used in this work are synthetic, the structural form of linear efficacy/off-target costs with pairwise cross-gene couplings which reflects established biological considerations~\cite{doench2016optimized}. By construction $Q_{ij} = 0$ for within-gene pairs, as such terms are identically zero under the one-hot constraint.

\subsection{The Greedy Trap}
A na\"ive biological heuristic selects the individually most potent candidate per gene, the one with highest on-target score. For our instance this yields $\{x_0, x_4, x_8\}$, ranked \emph{last} (64/64) among feasible solutions at cost $C = -0.75$, because these three candidates share the strongest positive pairwise couplings ($Q_{0,4} = +0.80$, $Q_{4,8} = +0.30$, $Q_{0,8} = +0.25$). The true combinatorial optimum is degenerate: two distinct selections, $\{x_1, x_5, x_9\}$ and $\{x_1, x_5, x_{10}\}$, both achieve $C = -2.350$. The optimum differs from the greedy pick at every gene, illustrating that single-gene optimization fails when pairwise cross-gene interactions are non-trivial. This motivates combinatorial optimization.

\subsection{Two Constraint-Handling Strategies}
We compare two formulations of the optimization (Fig.~\ref{fig:qubo_formulation}). The \emph{penalty} approach adds terms $P(\sum_{i \in S_k} x_i - 1)^2$ to $C(x)$, yielding an unconstrained QUBO whose ground state lies in the feasible subspace for sufficiently large $P$. The \emph{XY-mixer} approach enforces Eq.~\ref{eq:onehot} structurally: the cost Hamiltonian encodes only $C(x)$, and constraint preservation is delegated to the XY-mixer $H_M^{XY}$ (Eq.~\ref{eq:xy_mixer}) initialized from a feasible one-hot state. This distinction of soft penalty in the objective versus hard structural enforcement in the mixer is the central methodological axis of our study.
\section{Methods}
\label{sec:methods}

\begin{figure}[h]
\centering
\includegraphics[width=\columnwidth]{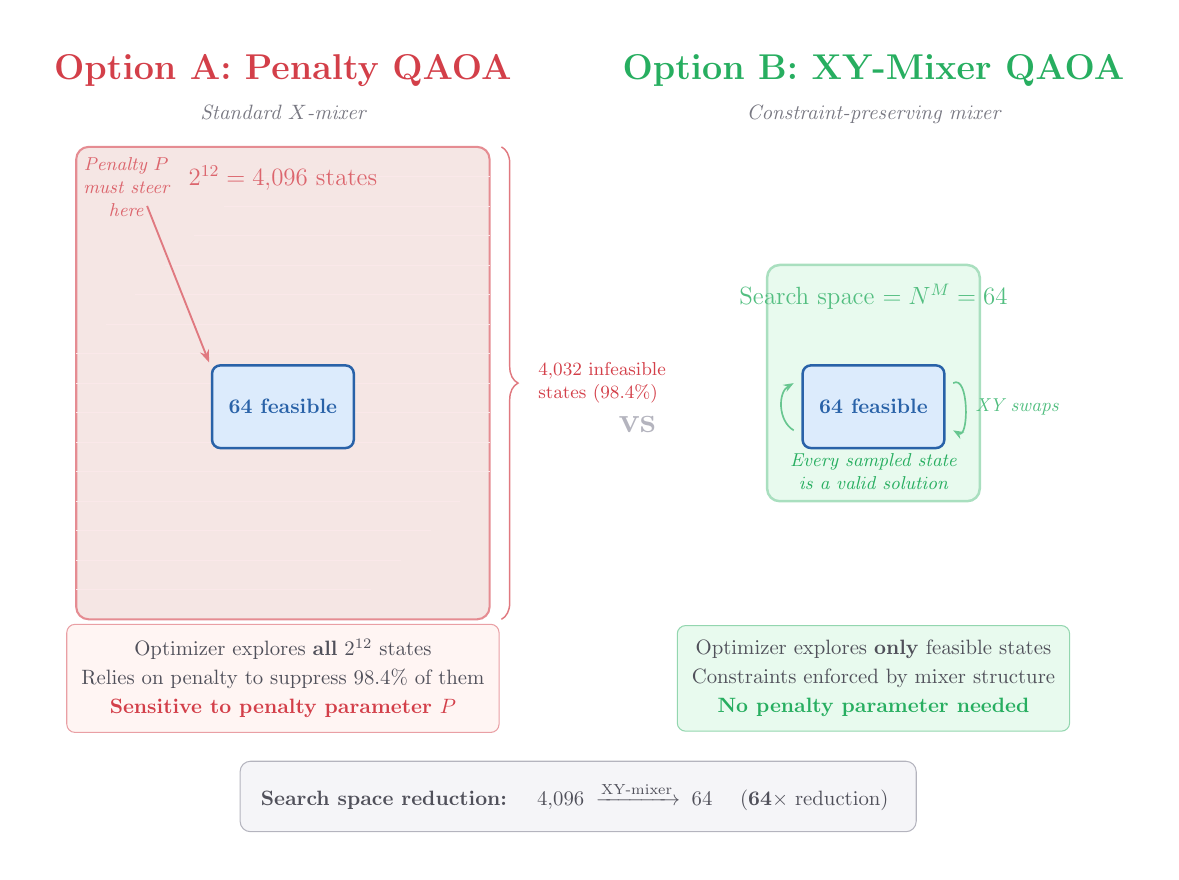}
\caption{The penalty approach searches all $2^{12}=4096$ bitstrings (98.4\% infeasible); the XY-mixer confines evolution to the 64 feasible states.}
\label{fig:search_space}
\end{figure}

\subsection{QUBO Construction and Penalty Variants}
We construct the constrained problem of Eq.~\ref{eq:cost}--\ref{eq:onehot} as a \texttt{QuadraticProgram} and convert it to QUBO form via Qiskit's \texttt{QuadraticProgramToQubo}, mapping to an Ising Hamiltonian $H_C$ through $x_i = (I - Z_i)/2$ .We evaluate three manual penalty coefficients $P \in \{2.0,\, 3.0,\, 5.0\}$ in simulation, of which $P \in \{2.0,\, 5.0\}$ are carried forward to hardware, alongside $P \approx 13.65$ returned by Qiskit's automatic penalty heuristic. For the XY-mixer variant we build a separate unconstrained \texttt{QuadraticProgram} encoding only $C(x)$; this yields a cost Hamiltonian with 60 Pauli terms, compared to 78 for every penalty variant (Fig.~\ref{fig:ham_complexity}).

\subsection{XY-Mixer and Initial State}
Within each gene group $S_k$ of 4 qubits, we implement the complete-graph XY-mixer of Eq.~\ref{eq:xy_mixer} as a sum of all 6 intra-group $(X_iX_j + Y_iY_j)/2$ terms, for a total of 36 Pauli terms across three groups~\cite{wang1904xy}. The initial state is $|\psi_0\rangle = X_0 X_4 X_8 |0\rangle^{\otimes 12}$, a one-hot configuration that selects the first candidate per gene. We emphasize that $\{x_0, x_4, x_8\}$ is the \emph{greedy-trap} configuration identified in \S\ref{sec:formulation}: the XY-mixer begins optimization from the worst-ranked feasible solution.

\subsection{QAOA Circuits and Classical Optimization}
For each variant and each depth $p \in \{1,2,3,4,5\}$, we construct a QAOA ansatz (Eq.~\ref{eq:qaoa}) with $2p$ variational angles. For penalty variants we use the default transverse-field mixer $H_M = \sum_i X_i$ with $|\psi_0\rangle = |+\rangle^{\otimes n}$; for the XY-mixer variant we substitute $H_M^{XY}$ and the one-hot initial state. Parameters are optimized by COBYLA (maximum 500 iterations) with 15 random restarts drawn uniformly from $[-\pi, \pi]^{2p}$. Best-energy parameters from this outer loop define the circuit submitted to hardware. For classical reference we include brute-force enumeration of the 64 feasible selections and a constraint-preserving simulated annealing solver (geometric schedule, 10{,}000 iterations, 20 restarts), both of which recover the optimum in milliseconds.

\begin{figure}[h]
\centering
\includegraphics[width=0.9\columnwidth]{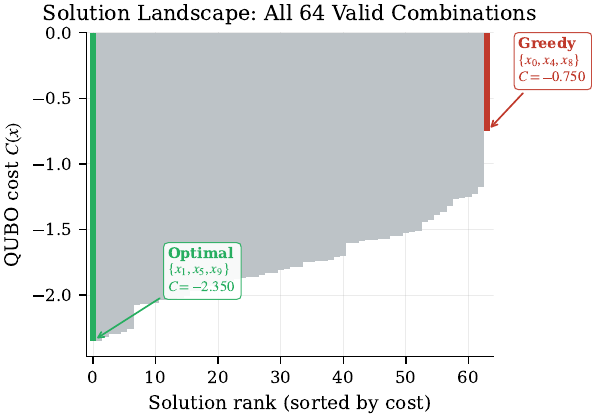}
\caption{All 64 feasible solutions ranked by cost. The combinatorial optimum $\{x_1, x_5, x_9\}$ (green, $C{=}-2.350$) and the greedy-trap configuration $\{x_0, x_4, x_8\}$ (red, $C{=}-0.750$) bracket the landscape. A second selection $\{x_1, x_5, x_{10}\}$ is cost-degenerate with the optimum.}
\label{fig:brute_force}
\end{figure}

\begin{figure*}[t]
\centering
\includegraphics[width=\textwidth]{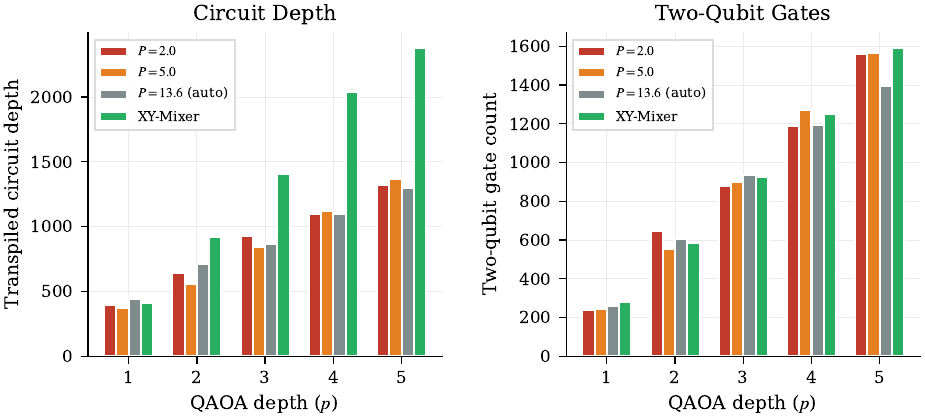}
\caption{Transpiled circuit depth and two-qubit gate count for each QAOA variant as a function of depth $p$, targeting \texttt{ibm\_kingston} (Heron r2) at optimization level 3.}
\label{fig:transpiled_metrics}
\end{figure*}

\subsection{Hardware Execution on ibm\_kingston}
Bound circuits are transpiled at optimization level 3 for IBM \texttt{ibm\_kingston} (Heron r2 processor). Fig.~\ref{fig:transpiled_metrics} reports transpiled circuit depth and two-qubit gate count per variant and depth. Jobs are submitted via the Qiskit Runtime V2 primitives. Energy measurements use \texttt{EstimatorV2}; bitstring sampling uses \texttt{SamplerV2}. We evaluate two mitigation settings:
\begin{itemize}[leftmargin=*,topsep=2pt,itemsep=1pt]
    \item \textbf{res=0}: no mitigation (raw).
    \item \textbf{res=2}: measurement twirling, dynamical decoupling with the XY4 sequence~\cite{viola1999dynamical,ezzell2023dynamical,ji2024synergistic}, and zero-noise extrapolation~\cite{temme2017error,li2017efficient,giurgica2020digital}.
\end{itemize}
Zero-noise extrapolation is supported by \texttt{EstimatorV2} but not \texttt{SamplerV2}; consequently sampler jobs at res=2 receive measurement twirling and DD but no ZNE. Each job uses 4{,}096 shots.

\subsection{Experimental Design}
We run each of the four variants at each depth $p \in \{1,\ldots,5\}$ under both resilience levels via \texttt{EstimatorV2} (40 energy jobs). Bitstring-level diagnostics (feasibility rate, sampling probability of the optimum) are obtained via \texttt{SamplerV2} for the $P=2.0$ and XY-mixer variants only (20 sampler jobs). To quantify run-to-run variation on hardware, we submit three additional res=2 repetitions per depth for the XY-mixer (15 jobs), yielding four hardware measurements per depth (one from the main sweep, three additional repetitions) for confidence-band estimation. The total experimental budget is 75 jobs.

\subsection{Metrics}
We report four quantities. The \emph{Ising energy} $\langle H_C \rangle$ measures the expectation of the cost Hamiltonian as executed (including the penalty term for penalty variants, excluding it for XY-mixer). The \emph{sim--HW gap} is the difference $\langle H_C \rangle_{\text{HW}} - \langle H_C \rangle_{\text{sim}}$ at matched parameters; positive gaps indicate noise-induced energy increase, while negative gaps indicate mitigation overcorrection. The \emph{feasibility rate} is the fraction of sampled bitstrings satisfying Eq.~\ref{eq:onehot}. The \emph{probability of the optimum} $P(\text{opt})$ is the combined sampling probability of the two cost-degenerate optima $\{x_1, x_5, x_9\}$ and $\{x_1, x_5, x_{10}\}$. We do not report approximation ratio: at this problem size and shot count, all methods trivially sample a bitstring achieving cost $-2.350$, rendering AR non-discriminative.

\section{Simulation Results}
\label{sec:sim_results}

\subsection{Classical Baselines}
Brute-force enumeration over the 64 feasible selections identifies a degenerate optimum at cost $-2.350$, and locates the greedy pick $\{x_0, x_4, x_8\}$ at rank 64 with cost $-0.750$ (Fig.~\ref{fig:brute_force}). A constraint-preserving simulated annealing solver recovers the optimum in all 20 independent runs within seconds, confirming that the problem is classically trivial at this size.

\subsection{Probability of the Optimum}
\begin{figure*}[t]
\centering
\begin{subfigure}[b]{0.46\textwidth}
    \centering
    \includegraphics[width=\textwidth]{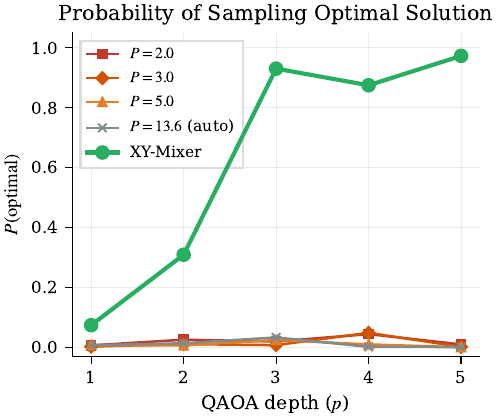}
    \caption{Probability of sampling a cost-optimal solution vs.\ QAOA depth. The XY-mixer exceeds $95\%$ from $p{=}3$; all penalty variants stay below $6\%$ at every depth.}
    \label{fig:p_opt}
\end{subfigure}
\hfill
\begin{subfigure}[b]{0.46\textwidth}
    \centering
    \includegraphics[width=\textwidth]{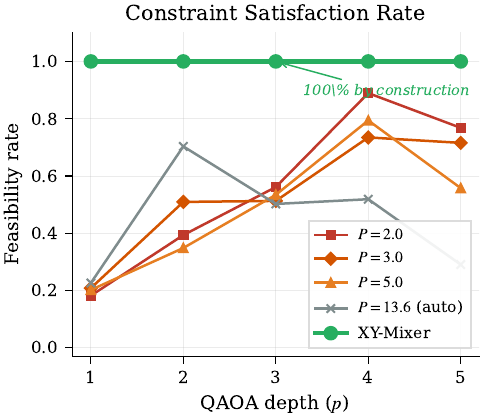}
    \caption{Feasibility rate vs.\ depth. The XY-mixer is pinned at $100\%$ by construction; penalty variants vary erratically between $18\%$ and $85\%$.}
    \label{fig:feas}
\end{subfigure}

\vspace{0.5em}

\begin{subfigure}[b]{0.46\textwidth}
    \centering
    \includegraphics[width=\textwidth]{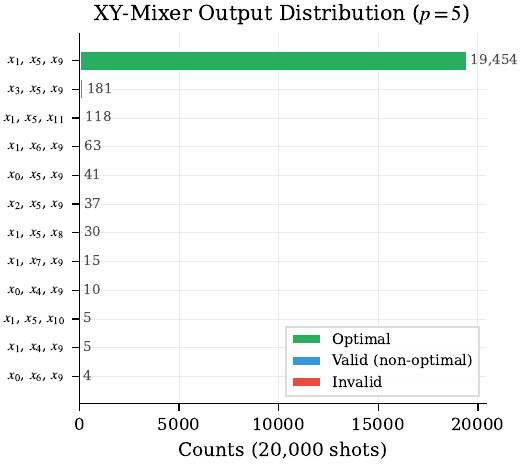}
    \caption{XY-mixer output distribution at its best-performing depth. Green: the two cost-degenerate optima. Blue: other feasible solutions. No infeasible bitstrings are sampled.}
    \label{fig:bitstrings}
\end{subfigure}
\hfill
\begin{subfigure}[b]{0.46\textwidth}
    \centering
    \includegraphics[width=\textwidth]{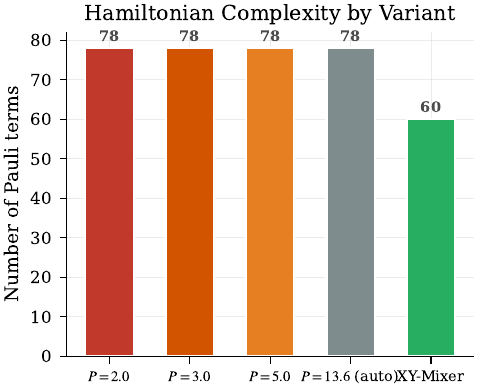}
    \caption{Number of Pauli terms in the cost Hamiltonian. The XY-mixer variant encodes only the objective $C(x)$ (60 terms), while all penalty variants include additional quadratic penalty terms (78 terms each).}
    \label{fig:ham_complexity}
\end{subfigure}

\caption{Simulation results for the four QAOA variants across depths $p=1$--$5$.}
\label{fig:sim_results}
\end{figure*}

Fig.~\ref{fig:p_opt} reports $P(\text{opt})$ as a function of QAOA depth for the four variants. The XY-mixer rises from $7.7\%$ at $p{=}1$ to $95.2\%$ at $p{=}3$ and plateaus thereafter, remaining above $94\%$ through $p{=}5$. All three penalty variants remain below $5.5\%$ at every depth, with $P{=}2.0$ peaking at $5.3\%$ ($p{=}4$), $P{=}5.0$ at $3.0\%$ ($p{=}2$), and $P{=}13.65$ at $1.7\%$ ($p{=}4$). The gap between the XY-mixer and the best penalty variant exceeds a factor of eighteen at the best-performing depths.

\subsection{Non-monotonicity and Landscape Pathology}
Penalty variants exhibit non-monotonic $P(\text{opt})$ and energy as a function of depth, in particular, for $P{=}5.0$ the optimized energy at $p{=}5$ ($-23.86$) is worse than at $p{=}4$ ($-25.67$), and similarly for $P{=}13.65$ ($-64.58$ vs $-70.76$). Because the depth-$p$ ansatz class strictly contains depth-$(p{-}1)$ as a subset, this regression is a signature of optimization difficulty: large penalty coefficients produce landscapes that the COBYLA optimizer fails to navigate as dimensionality grows. The XY-mixer shows only mild depth-to-depth variation attributable to restart variance.

\subsection{Feasibility}
The XY-mixer maintains $100\%$ feasibility at every depth by construction (Fig.~\ref{fig:feas}). Penalty variants produce feasibility rates ranging from $18\%$ to $85\%$ across the sweep, with no systematic trend in depth. Increasing $P$ does not monotonically improve feasibility: at $p{=}5$, feasibility falls from $84.9\%$ ($P{=}2.0$) to $41.8\%$ ($P{=}13.65$), contradicting the intuition that larger penalties more strongly enforce the constraint. This is a known consequence of penalty-induced landscape distortion~\cite{roch2023effect}: as $P$ grows, the cost Hamiltonian's spectral range widens and its optimization landscape worsens.


\subsection{Output Distribution Concentration}
Fig.~\ref{fig:bitstrings} shows the sampled bitstring distribution from the XY-mixer at its best-performing depth. The two degenerate optima together account for $95.2\%$ of 4{,}096 shots; the remaining probability mass is distributed among near-optimal feasible solutions ranked 3--10 by cost. No infeasible bitstrings appear, as expected. This concentration pattern contrasts sharply with the penalty variants, whose samples spread across both infeasible configurations and suboptimal feasible ones.


\subsection{Remark on Approximation Ratio}
At this problem size, the approximation ratio is uninformative: with 4{,}096 shots every variant samples at least one cost-$-2.350$ bitstring by chance, giving AR $\approx 1.000$ for all methods. The metrics above $P(\text{opt})$ and feasibility discriminate method quality in ways that AR cannot at this scale.
\section{Hardware Results}
\label{sec:hw_results}

\begin{figure*}[t]
\centering
\begin{subfigure}[b]{0.46\textwidth}
    \centering
    \includegraphics[width=\textwidth]{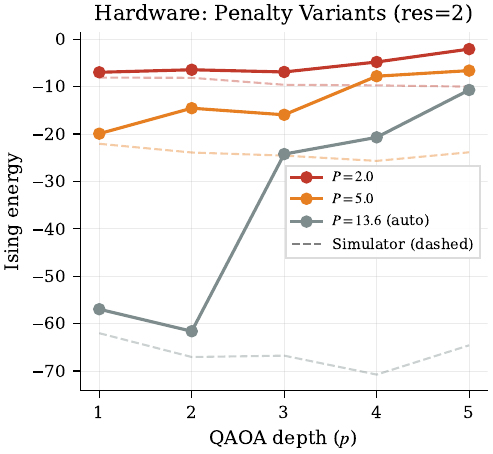}
    \caption{Penalty variants at res=2 (solid) versus simulator reference (dashed).}
    \label{fig:hw_penalty}
\end{subfigure}
\hfill
\begin{subfigure}[b]{0.46\textwidth}
    \centering
    \includegraphics[width=\textwidth]{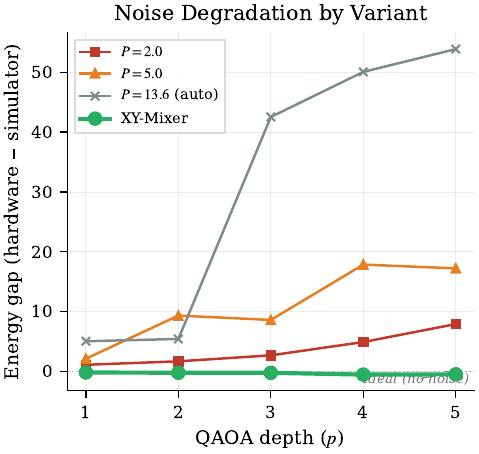}
    \caption{Sim--HW energy gap by variant at res=2. The $P{=}13.65$ gap reaches $+53.9$ at $p{=}5$; the XY-mixer remains near zero.}
    \label{fig:hw_gap}
\end{subfigure}

\vspace{0.5em}

\begin{subfigure}[b]{0.46\textwidth}
    \centering
    \includegraphics[width=\textwidth]{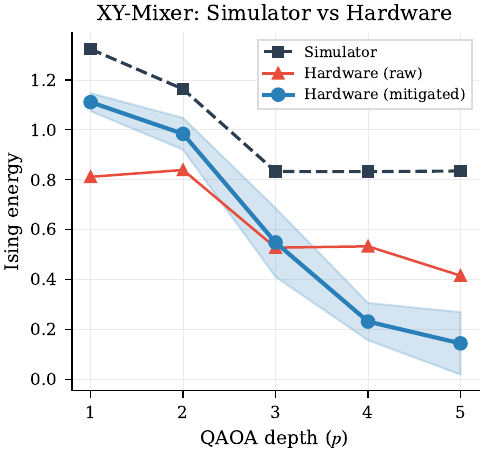}
    \caption{XY-mixer simulator vs.\ hardware. Shaded band: $\pm$std over four res=2 repetitions per depth.}
    \label{fig:hw_xy_energy}
\end{subfigure}
\hfill
\begin{subfigure}[b]{0.46\textwidth}
    \centering
    \includegraphics[width=\textwidth]{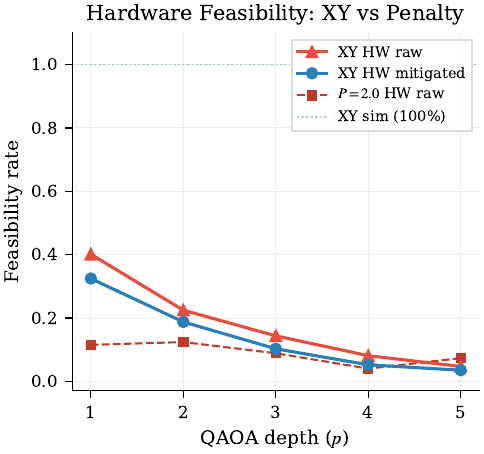}
    \caption{Hardware feasibility rate vs.\ depth. The XY-mixer's structural guarantee degrades under gate noise, decaying from $32.4\%$ ($p{=}1$) to $3.5\%$ ($p{=}5$); $P{=}2.0$ remains flat at $5$--$13\%$.}
    \label{fig:hw_feas}
\end{subfigure}

\caption{Hardware energy and feasibility measurements on \texttt{ibm\_kingston} (Heron r2, 4096 shots).}
\label{fig:hw_results_grid}
\end{figure*}

\subsection{Energy Gap Across Variants}
Fig.~\ref{fig:hw_gap} reports the hardware--simulator energy gap for all four variants at resilience level 2. Penalty variants exhibit monotonic growth of the gap with both depth and penalty coefficient: at $p{=}5$, the gap is $+7.9$ for $P{=}2.0$, $+17.2$ for $P{=}5.0$, and $+53.9$ for $P{=}13.65$. The $P{=}13.65$ variant exhibits a near-discontinuity between $p{=}2$ ($+5.4$) and $p{=}3$ ($+42.5$), consistent with the cost Hamiltonian's operator norm amplifying residual noise once circuit depth crosses a threshold. The XY-mixer, by contrast, maintains a gap of $-0.2$ to $-0.8$ across all depths---visually indistinguishable from zero on the scale of Fig.~\ref{fig:hw_gap}.

\subsection{XY-Mixer: Sim vs Hardware}
Fig.~\ref{fig:hw_xy_energy} overlays simulator, raw hardware, and error-mitigated hardware energies for the XY-mixer. The confidence band around the mitigated curve is the standard deviation across four independent res=2 measurements per depth. At shallow depth ($p{=}1,2$) the mitigated estimate lies between raw and simulator, consistent with partial noise correction. At $p{\geq}3$ the mitigated curve drops \emph{below} both raw hardware and simulator, producing the small negative gaps noted above. This is a characteristic signature of zero-noise extrapolation overcorrection under circuit-depth noise that violates ZNE's linear-in-noise-rate assumption~\cite{Mohammadipour2025Direct, majumdar2023best}. We report this faithfully rather than masking it: the mitigated values are not more accurate than the raw values at large $p$ for this variant.


\subsection{Sampling-Level Metrics}
Sampler jobs ran for the $P{=}2.0$ and XY-mixer variants only. The XY-mixer hardware $P(\text{opt})$ at res=2 peaks at $4.9\%$ ($p{=}3$), compared with $0.3\%$ for $P{=}2.0$ at its best depth---a 15-fold hardware-level advantage for structural constraint enforcement. Hardware feasibility for the XY-mixer decays monotonically with depth, from $32.4\%$ at $p{=}1$ to $3.5\%$ at $p{=}5$. The structural guarantee provided by $H_M^{XY}$ at the ideal-circuit level does not fully survive gate-level noise: two-qubit errors induce leakage out of the one-hot subspace. For $P{=}2.0$, hardware feasibility is lower and relatively flat ($5\%$--$13\%$ across depths). We note that at the deepest circuit ($p{=}5$), $P{=}2.0$'s feasibility ($5.4\%$) marginally exceeds the XY-mixer's ($3.5\%$)---an honest crossover we do not conceal.

\subsection{Bitstring Distribution on Hardware}
\begin{figure*}[h]
\centering
\includegraphics[width=\textwidth]{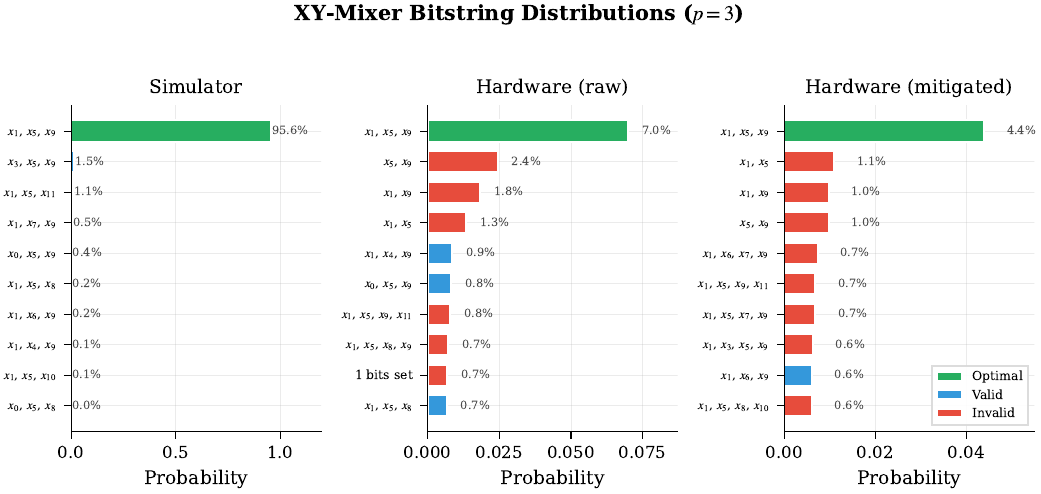}
\caption{XY-mixer output distributions at best hardware depth: simulator (left), raw hardware (center), mitigated hardware (right). Green bars: the two cost-degenerate optima.}
\label{fig:hw_bitstrings}
\end{figure*}

Fig.~\ref{fig:hw_bitstrings} compares the XY-mixer bitstring distribution at its best hardware depth across three settings: simulator, raw hardware, and mitigated hardware. The simulator concentrates $>94\%$ of probability on the two degenerate optima; on hardware, this concentration is diluted by noise, and the distribution spreads over a long tail of feasible and infeasible bitstrings. Measurement twirling and dynamical decoupling modestly sharpen the distribution relative to raw hardware but do not recover simulator-level concentration.

\subsection{Summary}
The XY-mixer retains a decisive advantage on hardware in energy accuracy (gap $|{-}0.8|$ vs $+53.9$ for the worst penalty variant at $p{=}5$) and in $P(\text{opt})$ at its best depth ($4.9\%$ vs $0.3\%$). However, both the structural feasibility guarantee and the sharpness of the output distribution degrade under noise, and ZNE overcorrection complicates mitigated-energy interpretation at large depth. These results provide an honest account of where constraint-preserving mixers succeed and where they partially fail on present-generation hardware.
\section{Discussion}
\label{sec:discussion}

\subsection{Structural vs Penalty Constraint Enforcement}
Our results separate two distinct failure modes of the penalty approach. First, penalty-induced \emph{landscape distortion}: the cost Hamiltonian's operator norm grows with $P$, widening the spectral range and producing optimization landscapes in which COBYLA fails even at simulator precision. The non-monotonic energy-vs-depth behavior observed for $P{=}5.0$ and $P{=}13.65$ in simulation (deeper ansatz class, worse optimized energy) is a direct manifestation~\cite{roch2023effect}. Second, penalty-induced \emph{hardware amplification}: the same operator-norm growth that distorts the landscape also amplifies residual gate noise in expectation-value estimation, producing the $+53.9$ energy gap at $P{=}13.65, p{=}5$. The XY-mixer avoids both pathologies by construction: $H_C$ encodes only the biological objective, keeping its operator norm bounded by problem-intrinsic quantities, while $H_M^{XY}$ confines evolution to the feasible subspace without penalty terms.

\subsection{The Limits of Structural Guarantees Under Noise}
The feasibility guarantee provided by $H_M^{XY}$ holds exactly at the level of ideal unitaries but is not preserved under gate-level error channels. Each two-qubit operation on \texttt{ibm\_kingston} can leak amplitude out of the Hamming-weight-one subspace, and such leakage accumulates with circuit depth. Our data quantify this: XY-mixer hardware feasibility decays from $32\%$ at $p{=}1$ to $3.5\%$ at $p{=}5$, and at the deepest circuit the $P{=}2.0$ penalty variant marginally exceeds it. This is a structural--empirical distinction worth stating plainly. Constraint-preserving mixers are an idealized-hardware construct; their advantage on real devices is quantitative, not categorical. Nonetheless, at hardware-relevant depths ($p{\leq}3$) where total error remains manageable, XY-mixer feasibility exceeds penalty feasibility by factors of 2--5, and $P(\text{opt})$ by factors of 15 or more.

\subsection{Error Mitigation as a Double-Edged Tool}
The ZNE behavior we observe on XY-mixer energies is consistent with recent analyses of extrapolation bias under non-ideal noise models~\cite{Mohammadipour2025Direct}. At shallow depth, ZNE moves the mitigated estimate toward the simulator value, correcting noise-induced energy drift. At larger depth, circuit noise structure diverges from the linear-in-noise-rate assumption underlying ZNE, and extrapolation overshoots the true value, producing small negative sim--HW gaps. We do not interpret these negative gaps as evidence of hardware outperforming simulation; they are an artifact of a mitigation protocol applied outside its regime of validity. A reviewer taking the raw hardware energy as a more conservative estimate arrives at the same qualitative conclusion: the XY-mixer closely tracks simulator performance on \texttt{ibm\_kingston}, with a different point estimate.

\subsection{Scope and Limitations}
We emphasize several limitations. The 12-qubit problem size is classically trivial; simulated annealing recovers the optimum in all 20 runs within seconds. We make no claim of quantum advantage, and none should be read into our results. Our gRNA scoring data is synthetic, parameterized to reproduce qualitative biological structure (linear efficacy/off-target costs, cross-gene pairwise couplings) rather than any specific experimental measurements; the methodological conclusions transfer to real scoring data, but the specific numerical optima do not. Our classical optimization protocol uses COBYLA with 15 random restarts, a simple baseline that is known to underperform more sophisticated strategies such as warm-starts and parameter concentration~\cite{bucher2026constrained}; with additional restarts the penalty variants might improve, though the structural XY-mixer advantage on hardware-noise-compatibility would remain unaffected. Finally, we evaluate only two resilience levels; a finer sweep of mitigation strategies could characterize the overcorrection crossover more precisely.

\subsection{Related Directions}
Constraint-preserving operator constructions beyond the XY-mixer have been explored, including Hamming Weight Operators in an adaptive framework~\cite{hao2026constraint} and warm-started XY-QAOA at larger scale on IBM hardware~\cite{bucher2026constrained}. Our contribution is complementary: we extend the structural-enforcement argument into the domain of multiplex CRISPR gRNA selection, provide systematic sim--HW comparison across a range of penalty coefficients, and report the honest breakdown of structural guarantees under real-hardware noise.
\section{Conclusion}
\label{sec:conclusion}

We presented COMET, a proof-of-concept application of constraint-preserving QAOA to multiplex CRISPR gRNA selection. Cast as a QUBO with one-hot constraints, the problem admits two distinct QAOA formulations: penalty-based with a transverse-field mixer, or penalty-free with the XY-mixer. On simulator, the XY-mixer exceeds $95\%$ probability of the optimum by depth $p{=}3$, while all three penalty variants we tested remain below $6\%$. On IBM's \texttt{ibm\_kingston} (Heron r2) processor, the XY-mixer's sim--hardware energy gap stays within $|0.8|$ across all depths, while the penalty variants' gap grows to $+53.9$ at the largest tested penalty coefficient. The structural feasibility guarantee provided by the XY-mixer does not fully survive gate-level noise (hardware feasibility decays from $32\%$ at $p{=}1$ to $3.5\%$ at $p{=}5$) but quantitative advantages over penalty variants persist at practical depths. We report a zero-noise-extrapolation overcorrection artifact at deep circuits, consistent with known limits of the mitigation protocol. Future work includes extension to larger gene panels with experimentally measured scoring data, warm-started initialization strategies, and finer characterization of the hardware-noise regime where structural enforcement ceases to dominate.

\bibliographystyle{IEEEtran}
\bibliography{references}

\vspace{12pt}
\end{document}